\documentclass[twocolumn,prabib,showpacs,amsmath,amssymb,epsfig,floatfix]{revtex4}
\usepackage{graphicx} 
\usepackage{color}
\begin{document}


\title{Correlations between mechanical, structural and dynamical properties of polymer nanocomposites}


\author{Aki Kutvonen$^1$, Giulia Rossi$^{1,*}$, Tapio Ala-Nissila$^{1,2}$}
\affiliation{$^1$Department of Applied Physics, Aalto University School of Science, P.O. Box 11100, FI-00076 Aalto, Espoo, Finland}
\affiliation{$^{2}$Department of Physics, Brown University, P.O. Box 1843, Providence, Rhode Island, 02912-1843, USA}
\email[]{giulia.rossi@inserm.fr}



\date{\today}

\begin{abstract}
We study the structural and dynamical mechanisms of reinforcement of a polymer nanocomposite via coarse-grained Molecular Dynamics simulations. In a regime of strong polymer-filler interactions, the stress at failure of the PNC is clearly correlated to structural quantities, such as the filler loading, the surface area of the polymer-filler interface and the network structure. Additionally, we find that small fillers, of the size of the polymer monomers, are the most effective at reinforcing the matrix, by surrounding the polymer chains and maximizing the number of strong polymer-filler interactions. Such a structural configuration is correlated to a dynamical feature, namely the minimization of the relative mobility of the fillers with respect to the polymer matrix.
\end{abstract}


\maketitle

\section{Introduction}

The properties of rubbers and plastics can be efficiently tuned by dispersing micro or nanosized additives (fillers) in the polymer matrix. Properties that can be influenced include rheology \cite{Mackay03}, mechanical \cite{Paul08}, electrical \cite{Kim08} and optical \cite{Caseri00} properties. In particular, the mechanical reinforcement induced by micro fillers in rubbery and glassy materials has been known empirically (and  exploited at an industrial level) for many years \cite{Balazs06}. In polymer nanocomposites (PNC), the size of the fillers is reduced down to the nanoscale and becomes comparable with the size of the polymers or even of their monomers. However, the exact physical mechanisms leading to mechanical reinforcement have proven elusive. Recently, intensive experimental \cite{Delcambre10,Wang10,Boucher11} and computational \cite{Riggleman07,Gersappe02,Gersappe11,LiuPCCP11,RigglemanJCP09,Goswami10,Toepperwein11,Starr11,Riggleman09} efforts have been devoted to answering this fundamental question with the aim to get to better design of reinforced materials. Many parameters play a role: the size and shape of the fillers, the polymer molecular mass and the presence of an entangled network, the strength of the physical and/or chemical interaction between the fillers and the polymers, the filler loading and dispersion.  \\

In this report, we focus on the issue of a polymer nanocomposite undergoing tensile tests, leading to mechanical failure. We model the PNC in a regime of strong polymer-filler interactions and optimal degree of dispersion \cite{Liu11}. Fillers are small if compared to the radius of gyration of the chains. Their diameter varies from being equal to the monomer diameter, to being twice as large. In this regime, the smallest fillers 
are known to achieve the highest degree of reinforcement \cite{LiuPCCP11,Gersappe02}. The ability of the small fillers to induce toughening has to be partly attributed to their large surface to volume ratio, which allows for the formation of a large number of strong polymer-filler bonds. But, as we show here, surface area alone is not enough to explain the filler size dependence of the PNC mechanical response. There are other factors that need to be considered for a full explanation. \\

Fillers affect the PNC dynamics, as well, with possible repercussions on the mechanical properties of the composite. Experimental results analyzing mobility in PNC have suggested that the diffusion processes in PNC are far more complex than in pure polymers. For example, non-monotonic trends of chain diffusion coefficients vs. loading have been very recently observed experimentally in PNC containing both carbon nanotubes and nanoparticles \cite{Mu09,Mu11}, but no microscopic explanation is currently available. For smaller fillers, Gersappe \cite{Gersappe02} proposed that the reinforcement mechanism is based on the dissipation of deformation energy offered by the high mobility of the nanoparticles in the polymer matrix.\\

In order to clarify which are the structural and dynamical contributions to reinforcement, we study here the correlation between mechanical toughening, structural configuration and dynamics of the PNC. We find that the mechanical reinforcement, quantified by the stress at failure, is positively correlated with the creation of a network of numerous and strong polymer-filler bonds. Small fillers, whose size is comparable to the size of the polymer monomers, are the most effective at creating such a network. Concerning mobility, the toughening is  correlated to the minimization of the diffusion of the fillers with respect to the polymer matrix. We do not find any evidence for dissipation effects as proposed by Gersappe \cite{Gersappe02}.  \\



\section{Model and simulation methods}

\subsection*{Model system and interactions}
Our results are based on Molecular Dynamics (MD) simulations of a polymer melt containing 64 chains, each one composed by 64 beads, and a variable number of spherical filler particles. Intra-chain bonds are modeled by a 
harmonic potential, $E_{harm}(r)=\frac{1}{2}k_{h}(r-r_{h}^{0})^{2}$. $r$ is the distance between the monomers, $k_{h}$ is the elastic constant ($k_{h}=4000$ kJ mol$^{-1}$ nm$^{-2}$) and $r_{h}^{0}$ is the equilibrium length of the bond ($r_{h}^{0}=0.47 $ nm). The elastic constant was set at the minimum value preventing chains from crossing each other \cite{Nikunen07}. \\
Non-bonded interactions are modeled by Lennard-Jones potentials, $V_{LJ}(r)=4\epsilon ((\sigma/r)^{12}-(\sigma/r)^{6})$. In the model, the reference energy scale is provided by the polymer-polymer LJ interactions, with $\sigma=0.47 $ nm and $\epsilon=0.05$ eV. These values set the degree of coarse-graining of our model, where each bead can represent realistically a chemical moiety containing four carbon atoms \cite{Marrink07}. The filler-filler interaction is weaker than the polymer-polymer interaction $\epsilon$,  with $\epsilon_{ff}=\epsilon /5$. The polymer-filler interaction is stronger, with $\epsilon_{pf}=4\epsilon$. Three different sizes for the fillers were considered. Our length scale is set by the size of the chain monomer, $\sigma$.  The mass of the polymer beads has been set to the value of 56 amu, corresponding again to a (CH$_2$)$_4$ group and providing a density of 730 kg/m$^3$ (0.81 chains/$\sigma^3$ \footnote{The polymer chain length is below the entanglement length, as predicted by primitive path analysis \cite{Everaers04} at the density of 0.85 chains/$\sigma^3$} ) at 600 K and atmospheric pressure conditions. The small filler (SF) is as large as the chain monomer and has the same mass, while the medium and large fillers (MF and LF, respectively) have size 1.3$\sigma$ and 1.9$\sigma$, their masses scaling accordingly to their volumes. Values of $\sigma$ for polymer-filler Lennard-Jones interactions were chosen as the arithmetic average between the $\sigma$ of the polymer and that of the filler. The addition of fillers strongly interacting with the polymer chains has the effect to increase the density of the composite, whose value depends on the filler size and concentration. For example, for the sample containing 21\% SF, the density increases to 870 kg/m3. \\

\subsection*{Simulation protocols}

\textbf{System set-up.} Each of our initial, independent configurations was set up as following. First, we placed the colloidal nanoparticles at random positions within a large simulation box. Then, we placed the first monomer of a polymer chain, again choosing at random its position. We placed the rest of the monomers one after the other, along a random direction and at equilibrium distance from the previous one, avoiding overlappings. The procedure was then iterated for each of the 64 polymer chains.\\

\textbf{Equilibration.} MD simulations were performed by means of the GROMACS 4 simulation package \cite{Hess08}. For each composition and replica, we performed a 100 ns equilibration in the NpT ensemble (temperature controlled by a Nos\'{e}-Hoover thermostat, pressure controlled by a Parrinello-Rahman barostat), at a temperature of 600 K and a pressure of 1 bar. In these thermodynamic conditions, the pure polymer matrix is in the liquid state, well above its glass transition temperature that we estimated, by means of a slow cooling simulation, to be 390 K. As shown in more detail in Section \ref{sec:results}, at the temperature of 600 K all the filler-containing systems are in the liquid state, as well. The relaxation time of the end-to-end vector of the polymer chains varies between 10$^3$ and 10$^5$ ps depending on the nanofiller size and loading, assuring for a proper relaxation of the chain conformations during our equilibration run. During equilibration, periodic boundary conditions were applied in all directions.

\begin{figure}[h!]
\includegraphics[width=0.48\textwidth]{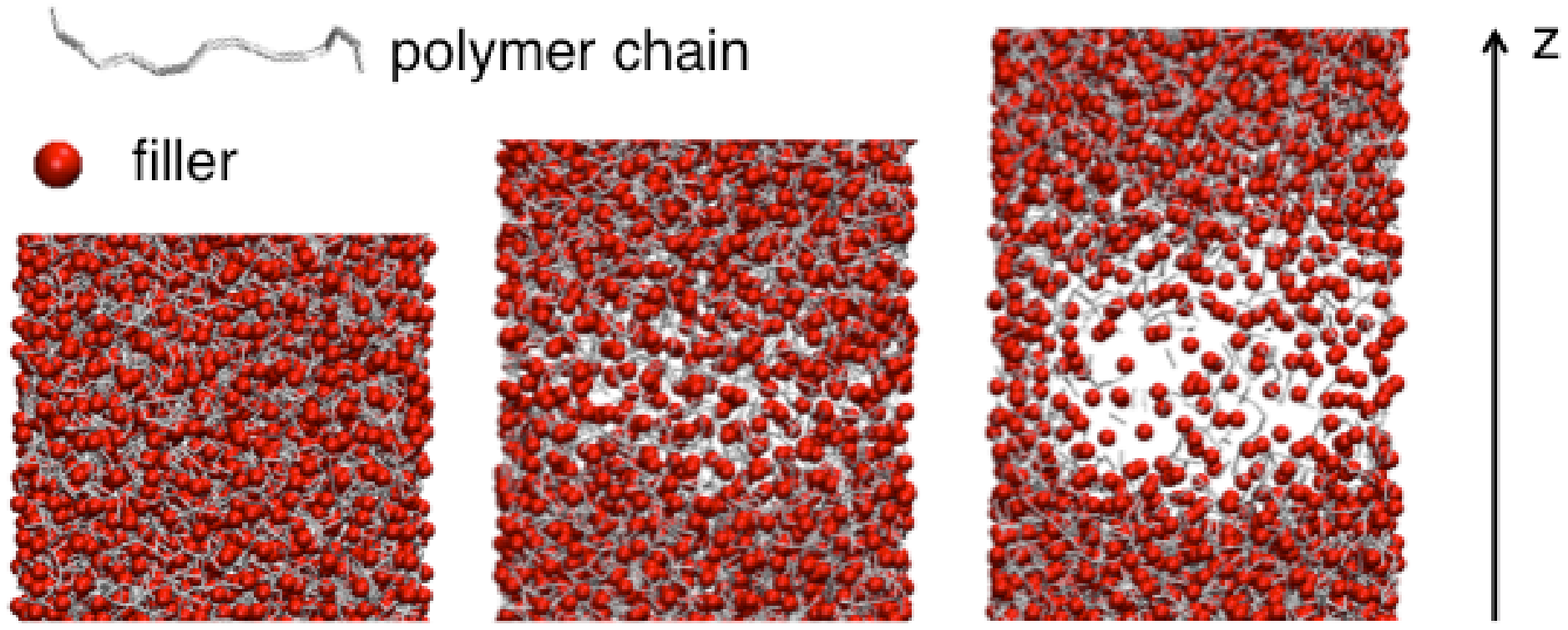}
\includegraphics[width=0.4\textwidth]{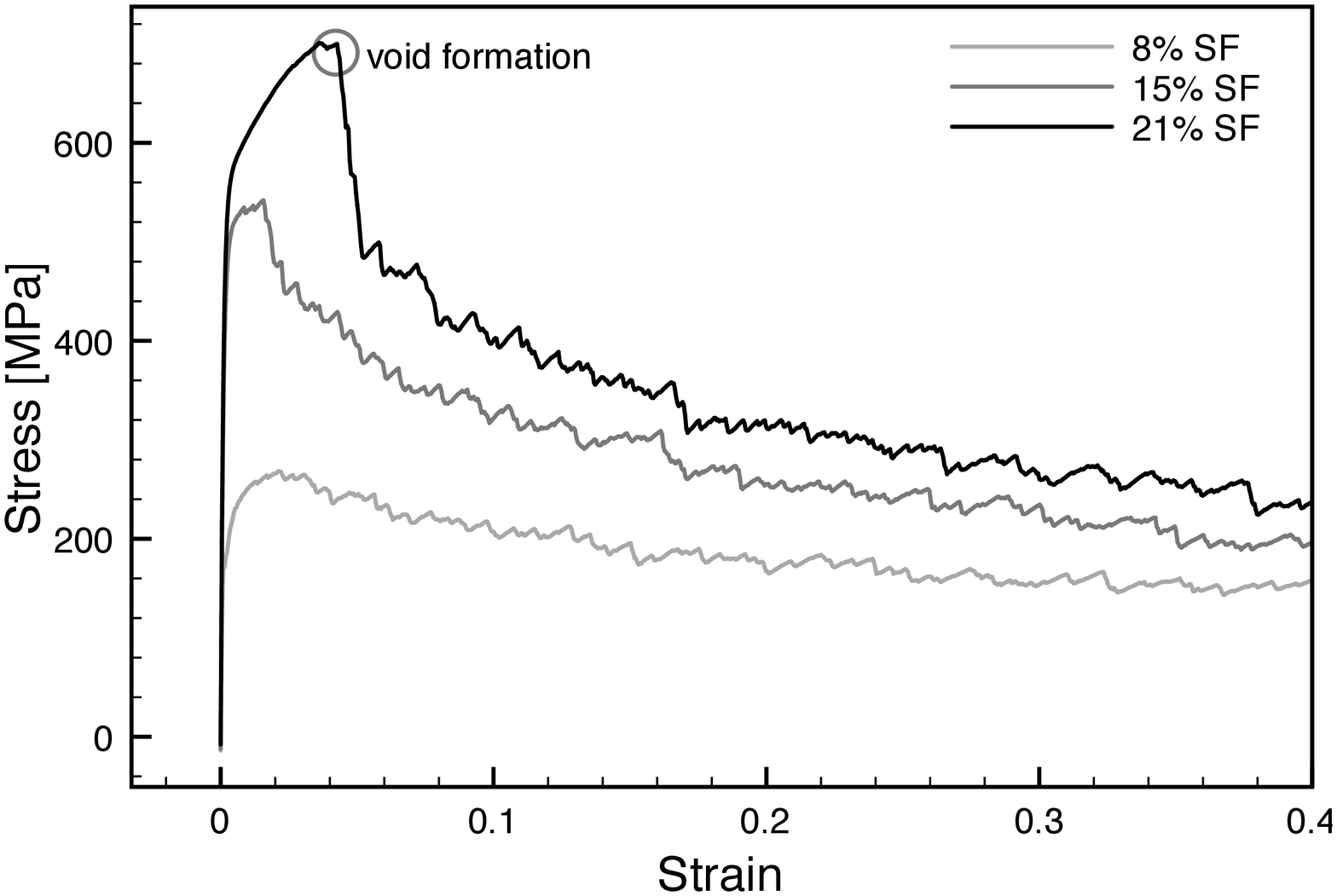}
\includegraphics[width=0.4\textwidth]{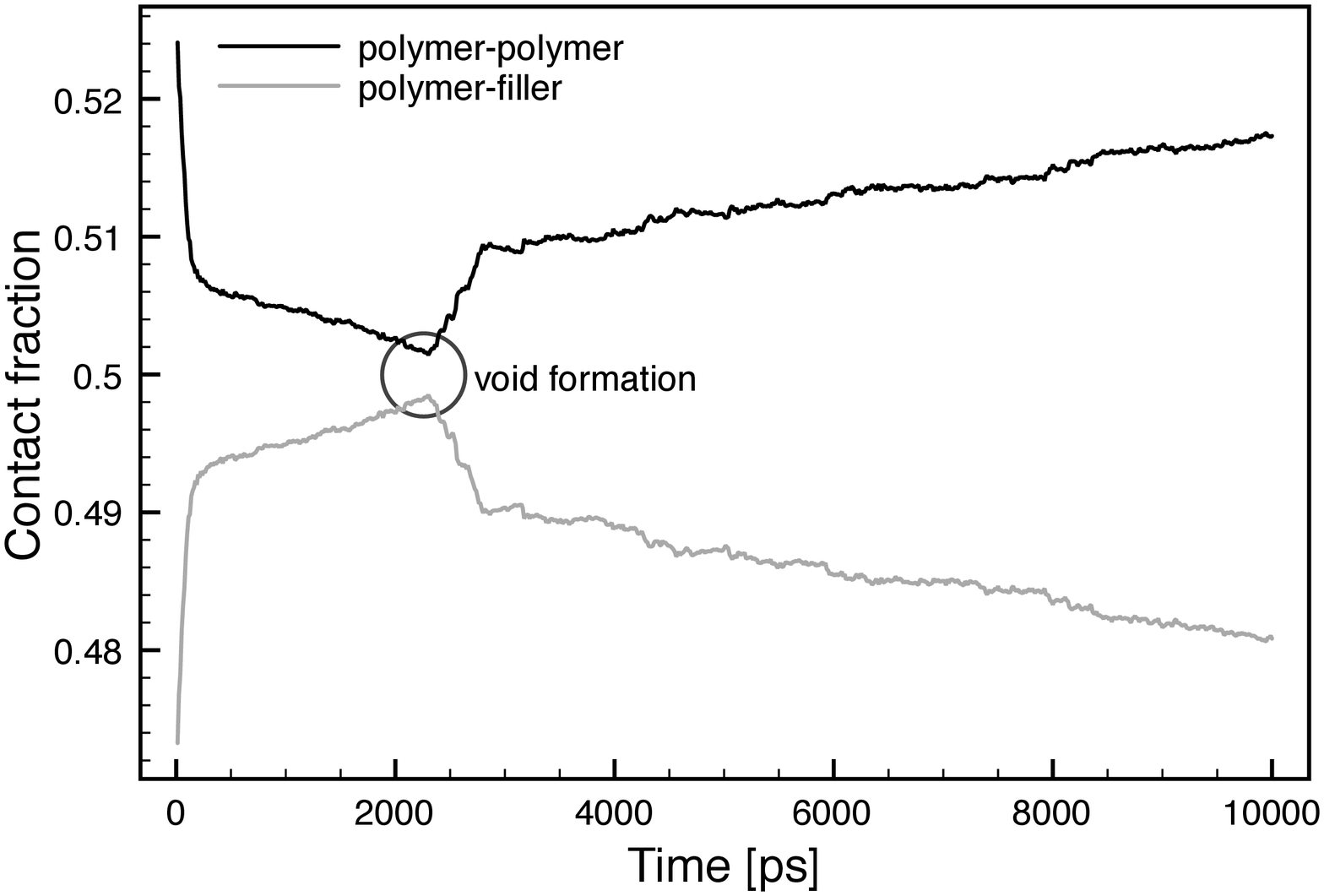}
\caption{\label{fig:snap} Top: snapshots from a tensile test (small fillers, 21\% mass loading). Chain bonds are represented by light grey sticks. Filler particles are shown as red beads. Middle: some examples of stress-strain curves as obtained for PNC loaded with 8\%, 15\% and 21\% SF. Each data set results from the average over 5 independent tensile tests. The grey circle on top of the stress-strain curve for the 21\% SF system indicates the onset of the formation of a single large void in the PNC matrix. Bottom: evolution of the fraction of polymer-polymer and polymer-filler contacts during the pulling of a PNC loaded with 21\% SF. Void formation is again indicated by a grey circle. The fraction of filler-filler contacts is negligible (10$^{-3}$, decreasing to 10$^{-4}$ before void formation). }
\end{figure}

\textbf{Tensile tests.} After equilibration, the thermostat and the barostat were turned off and the tensile test started. We performed our tensile tests by stretching the $z$ edge of the simulation box at constant velocity, while keeping the $x$ and $y$ at fixed length (see top panel of Fig.\ref{fig:snap}). Periodic boundary conditions were still applied in all directions. The $z$ edge of the simulation box was enlarged at the constant velocity of 
$v_p$=1.1 m/s (a value close to the one used in previous studies, as in \cite{Gersappe02}). A further set of tensile tests, not reported in this paper, was performed according to a second protocol, mimicking exactly the set-up used by Gersappe \cite{Gersappe02}. In this latter case, the PNC is confined between two sticky walls, which are pulled apart during the test. We remark that none of the results depends, at least qualitatively, on the choice of one these alternative protocols (about the role of boundary conditions in tensile test simulations, see also Rottler and Robbins \cite{Rottler03}). 

\begin{figure}[h!]
\includegraphics[width=0.45\textwidth]{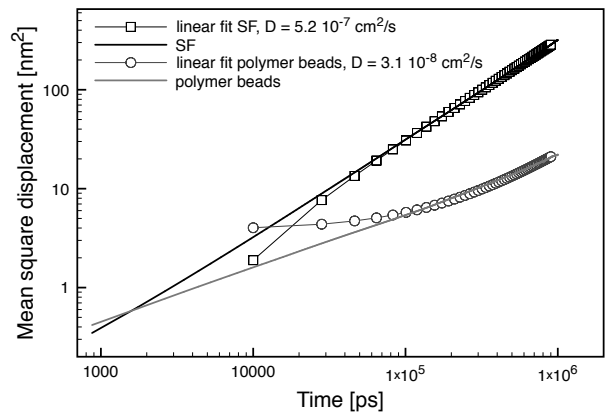}
\caption{\label{fig:msd} Mean square displacement of SF and chain monomers, at T = 600 K and atmospheric pressure. In the inset, the grey area is zoomed and the transient regime for chain monomers is shown.}
\end{figure}

\textbf{Measure of the diffusion coefficient.}
$D$ was calculated from the mean square displacement $\langle|\Delta R|^{2}\rangle = \langle |R(t)-R(0)|^{2} \rangle$ of the particles, averaged over time and over all particles. Data were recorded from 1 $\mu$s long runs in the NpT ensemble, at $T=600$ K and $p=1$ bar, under periodic boundary conditions, and $D$ was obtained via linear fit from the relation $6D=\lim_{t\rightarrow \infty}\langle |\Delta R|^{2}  \rangle/t$. As an example, Fig. \ref{fig:msd} shows the mean square displacement data for a PNC a 21\% mass loading of SF. While SF reach terminal diffusion very quickly, on the time scale of tens of ns, the chain monomers have a slower diffusion and the linear regime is reached after about 200 ns.

\section{Results}
\label{sec:results}

\subsection*{Tensile tests}
\begin{figure}[h!]
\includegraphics[width=0.4\textwidth]{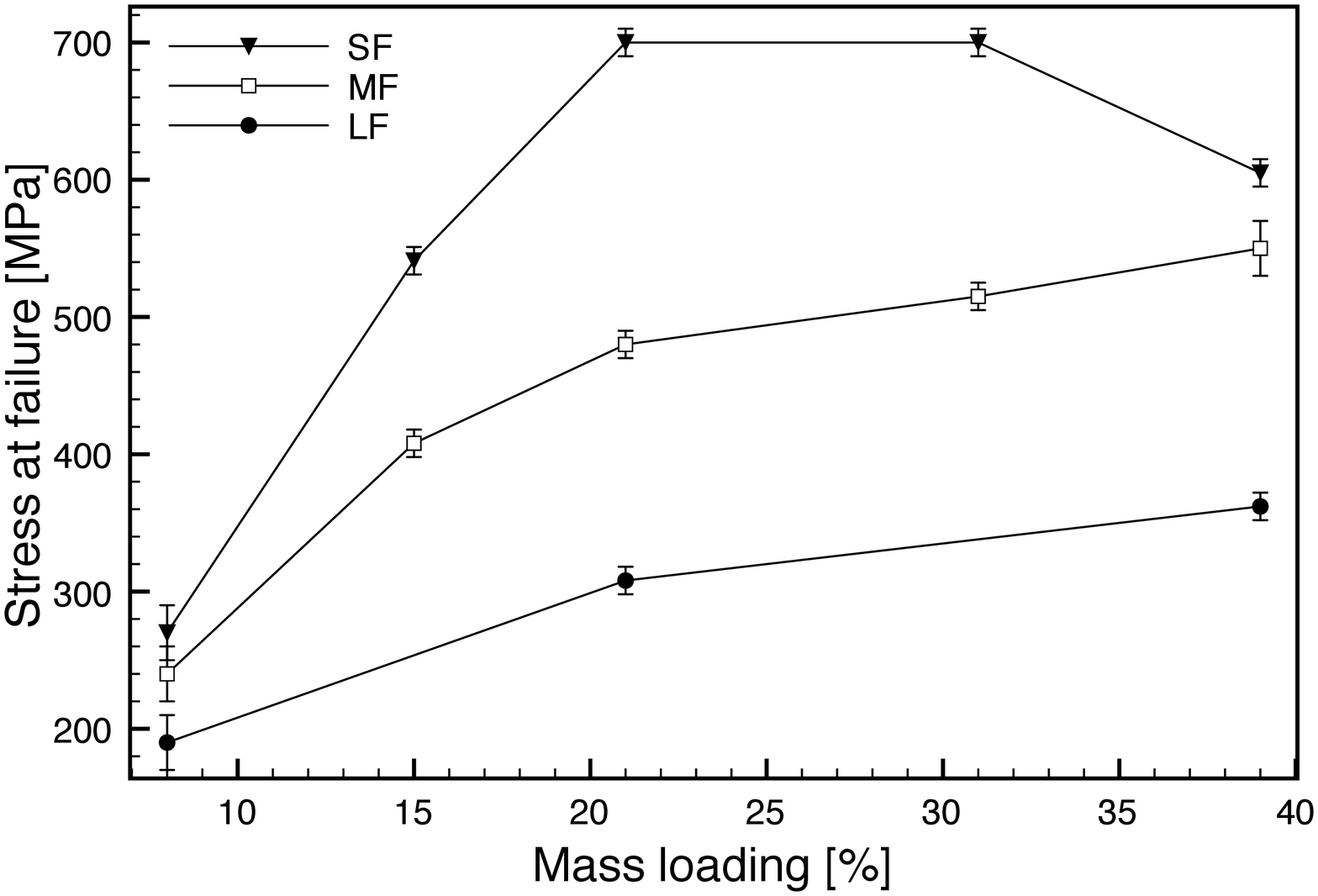}
\includegraphics[width=0.4\textwidth]{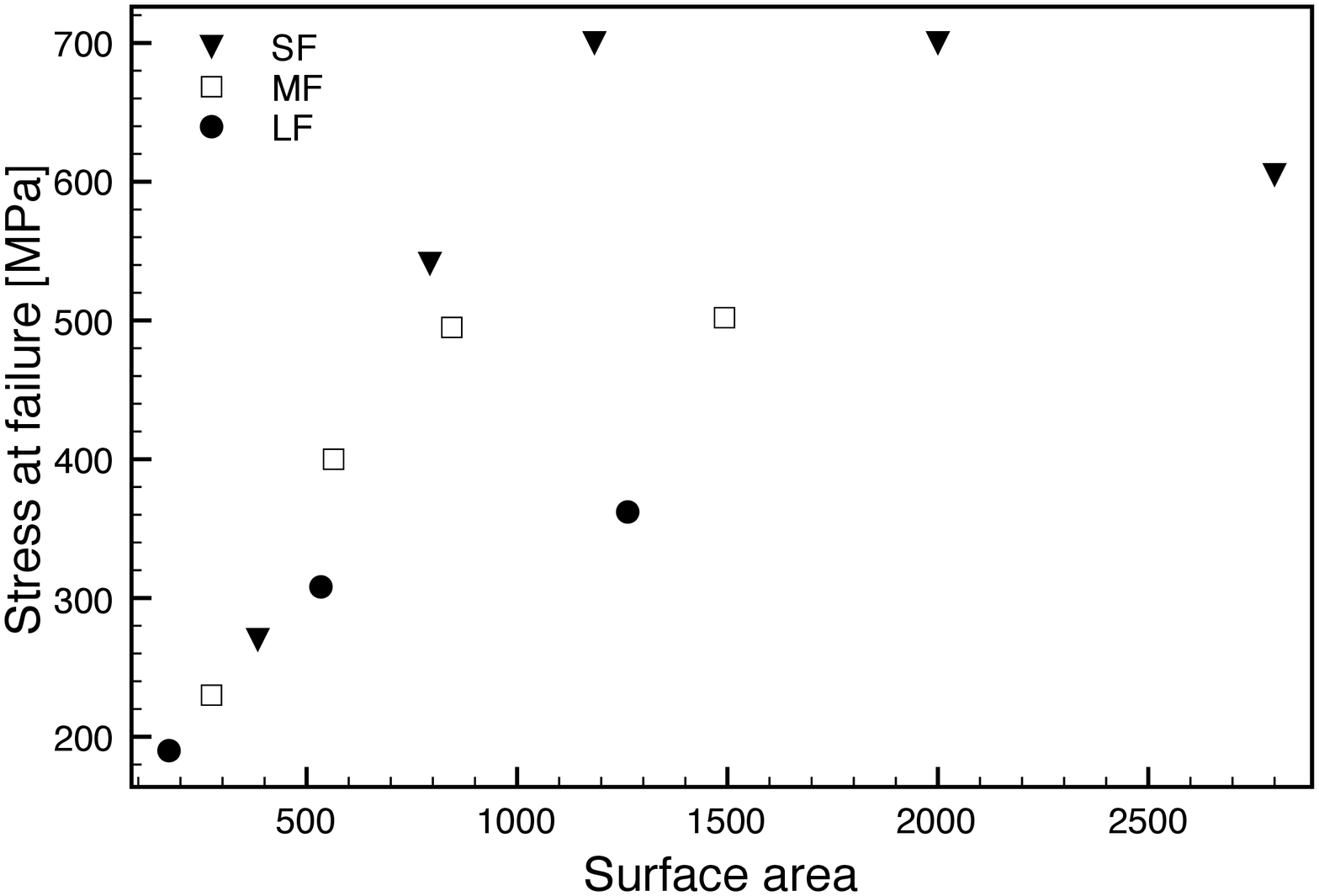}
\caption{\label{fig:stressstrain} Top: stress at failure as a function of filler mass loading for SF, MF and LF. Bottom: stress at failure as a function of filler total surface area for the same systems reported in the upper panel. Surface area is expressed in units of the SF surface area. }
\end{figure}

The top panel of Fig. \ref{fig:snap} depicts a typical evolution of a tensile test simulation, and its stress-strain profile (middle panel). After an initial homogeneous decrease of density during which the material's response is nearly elastic, the stress reaches a maximum (the stress at failure), the nanocomposite undergoes mechanical failure and a single, large void is formed in the matrix while the stress drops. Before void formation, the system reacts to pulling by forming as many strong polymer-filler contacts as possible. This is illustrated in the bottom panel of Fig. \ref{fig:snap}. A contact is formed any time two particles are found closer than 1.1$\times 2^{1/6} \sigma$, where $\sigma$ is the parameter of their mutual LJ interaction. In the Figure we plot contact fractions, namely the ratios between the polymer-polymer or polymer-filler contacts and the total number of contacts. Irrespective of the filler size, during pulling the fraction of contacts between the polymer chains and the fillers increases at the expense of polymer-polymer and filler-filler contacts. \\ 
Stress-strain curves allow to single out the stress at failure of the PNC matrix, here defined as the highest stress value achieved. The higher the stress at failure, the better the reinfocement offered by the nanofillers. As shown in Fig. \ref{fig:stressstrain}, small fillers provide the best reinforcement, achieving the largest stress at failure in the whole loading range considered. 

The availability of contact surface between the polymers and the fillers depend on the fillers' size, namely, the smaller the filler, the larger the contact surface area. Nevertheless, it has been shown in previous works such as in \cite{Gersappe02} that surface area is not the only cause for better reinforcement by the small fillers. In order to confirm this point, we plot the stress at failure as a function of the total surface area of the fillers, as shown in the bottom panel of Fig. \ref{fig:stressstrain}. At very low loadings the stress at failure is positively correlated with the filler surface area, but such a correlation quickly vanishes and it is completely lost at 21\% loading. In this regime, small fillers achieve the largest degree of reinforcement no matter what their total surface area is.\\


\subsection*{PNC dynamics}
\begin{figure}[h!]
\includegraphics[width=0.235\textwidth]{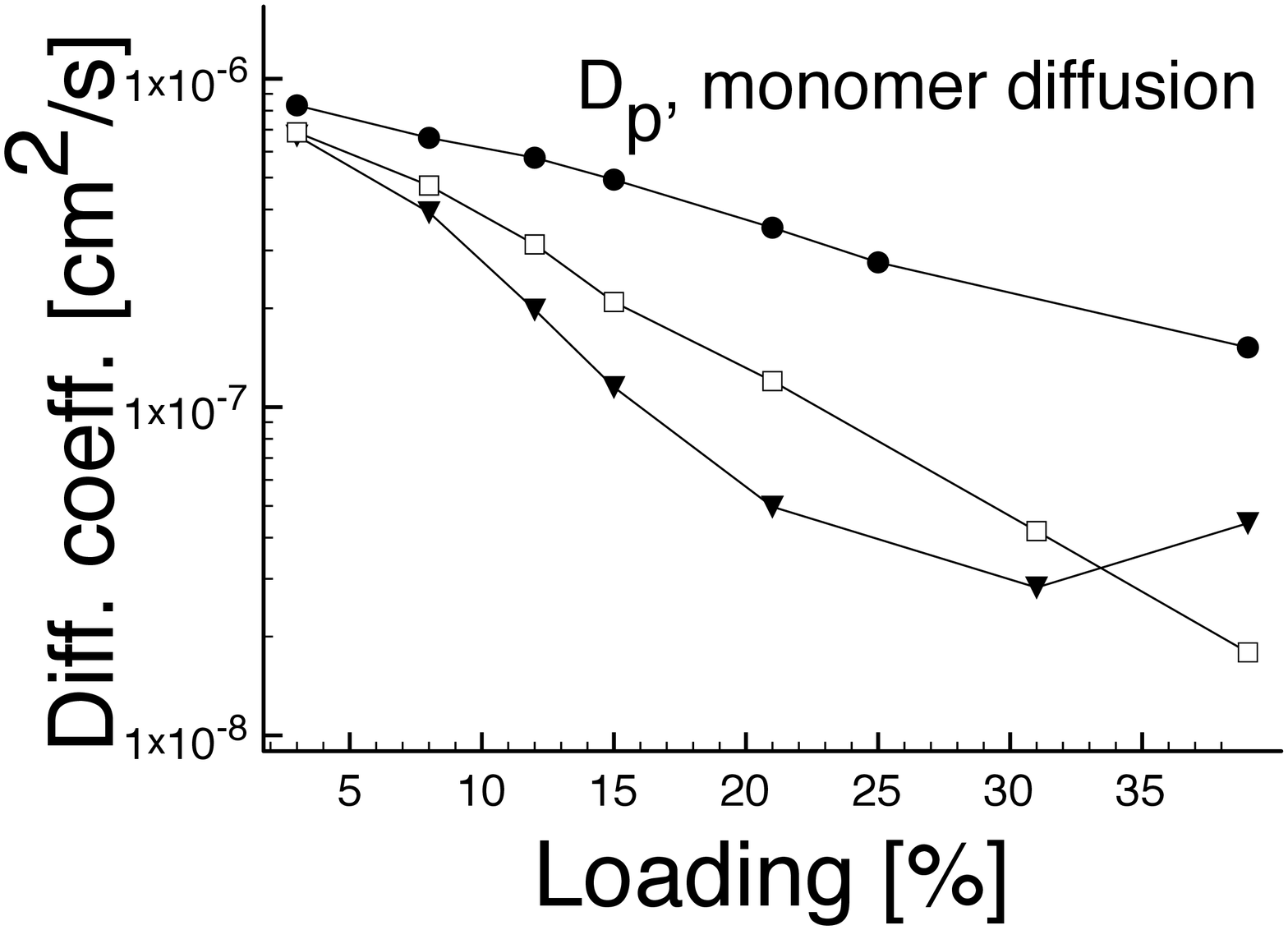}
\includegraphics[width=0.235\textwidth]{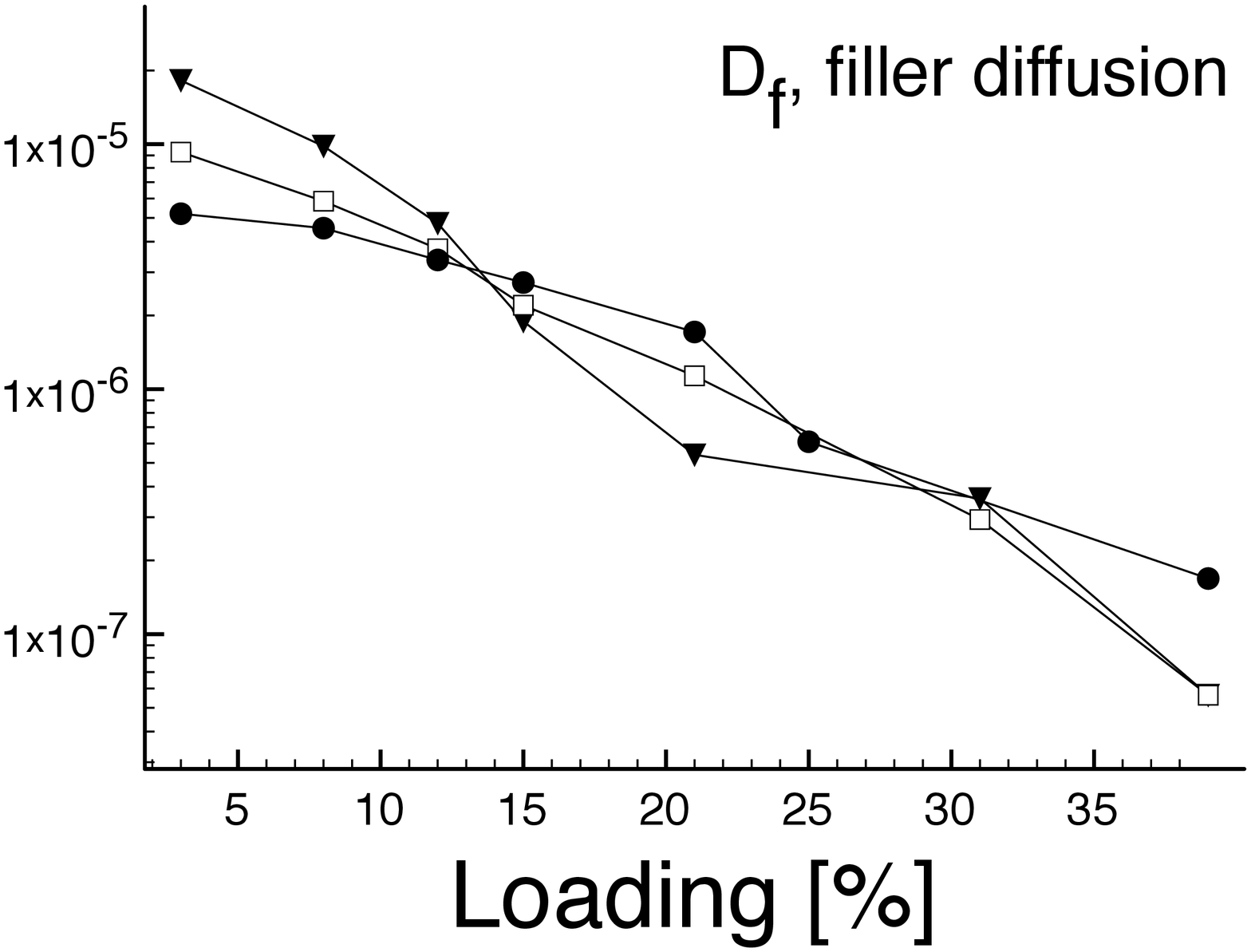}
\begin{center}
\includegraphics[width=0.48\textwidth]{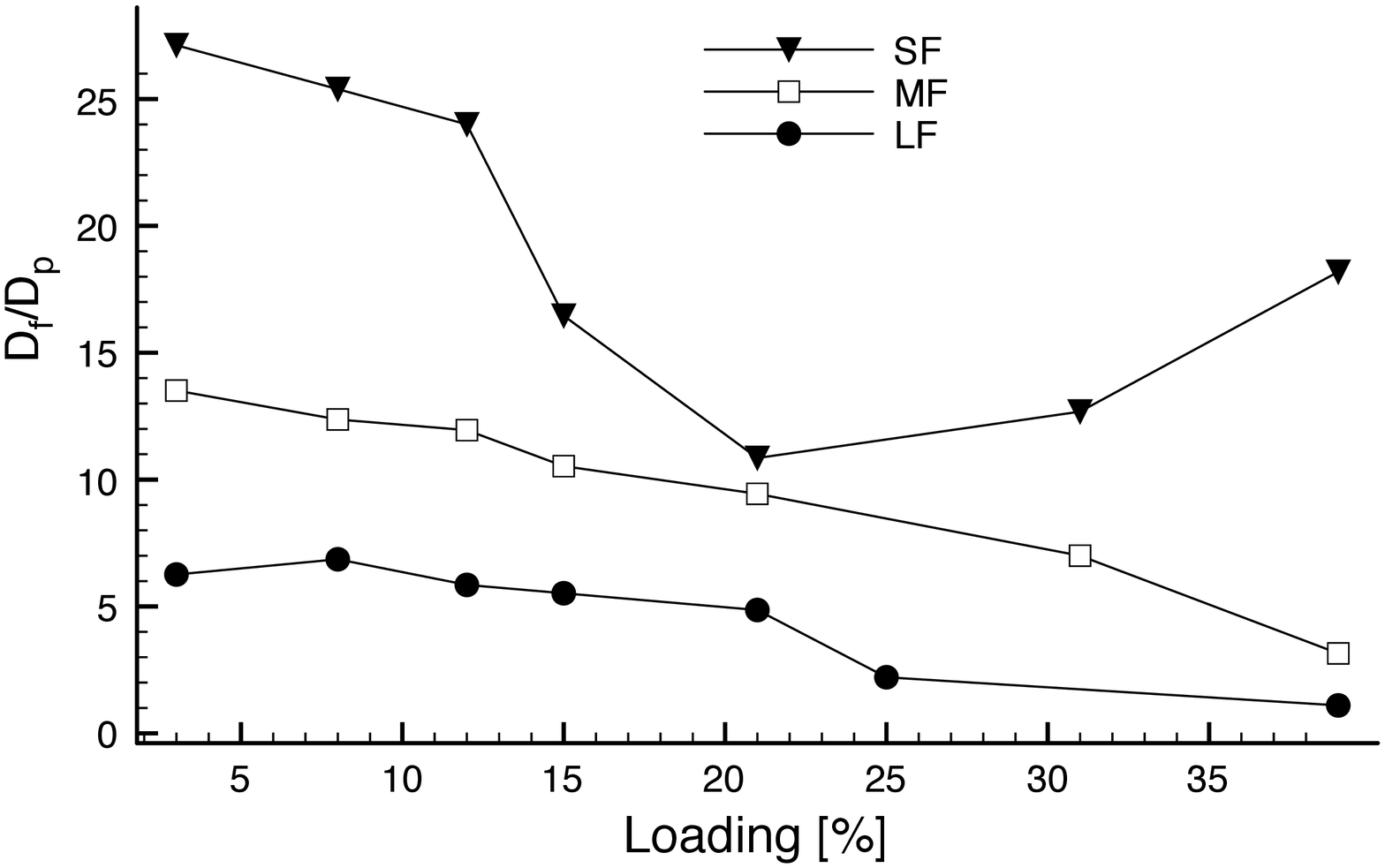}
\end{center}
\caption{\label{fig:diff} Top, left: diffusion coefficients of the polymer monomers as a function of filler loading and size. Top, right: diffusion coefficients of the fillers as a function of filler loading and size. Bottom: relative mobility, namely ratio between the filler and the polymer diffusion coefficients, as a function of loading and size. Lines are only guides for the eye.}
\end{figure}

\begin{figure}[h!]
\includegraphics[width=0.48\textwidth]{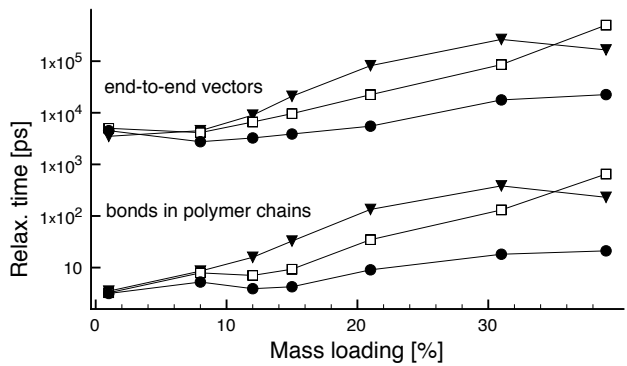}
\caption{\label{fig:trelax} Relaxation times of the polymer end-to-end vectors and of the polymer chain bonds in the PNC, as a function of filler loading and size.}
\end{figure}

In order to clarify whether the filler's mobility plays a role in determining such a dependence of the mechanical reinforcement on the filler size, we thus focused on the dynamics in our PNCs. We measured the diffusion coefficient, $D$, of the monomers and of the fillers as a function of loading, for the SF, MF and LF cases. \\ 
Let us start with the polymers. As shown in the top left panel of Fig. \ref{fig:diff}, their diffusion coefficient decreases nearly exponentially in the whole loading range considered, for the case of large and medium fillers dispersed in the matrix. An overall reduction of polymer diffusion vs. particle loading is indeed expected in a regime of strong polymer-filler interactions \cite{Desai05}. When coupled to small fillers, the monomer diffusion coefficient decreases nearly exponentially until 15\% loading, then reaches a local minimum at 31\% loading. Other dynamic indicators of the mobility of polymers in the PNC are the relaxation times derived from the polymer rotational autocorrelation functions. We looked at the rotational autocorrelation function, $C_{b}(t)$, of the end-to-end vector $\vec{b}$ defined as $C_{b}(t)=\langle P_{2}[\hat{b}(t)\cdot\hat{b}(0)] \rangle$, $P_2$ being the second Legendre polynomial and $\hat{b}$ a unitary vector directed as $\vec{b}$. We define the relaxation time as the time $\tau$ at which $C_{b}(t)$ reaches a value of 0.3. Relaxation time data are reported in Fig. \ref{fig:trelax}. The relaxation time of the end-to-end vector in the systems containing the SF has a maximum at composition 31\%. For MF- and LF-containing systems, the relaxation time increases monotonically with loading in the loading range considered. The relaxation times of the monomer-monomer bonds, analogously defined, show the same trend. \\ 
We can now analyze the fillers' mobility. As shown in the top right panel of Fig. \ref{fig:diff}, despite the fact that small fillers provide the best reinforcement effects in all the loading range considered, they do not always diffuse faster than the other particles. Between 12 and 15\% loading, the diffusion coefficients of the SF, MF and LF invert their order, the small fillers becoming the slowest in the PNC. \\
It has been suggested, by means of MD simulations, that the small fillers, namely those with the largest mobility relative to the matrix, should always be better at toughening the PNC \cite{Gersappe02} through the dissipation of deformation energy. This is questionable, as the time scale of the mechanical deformation in simulations is much shorter than the diffusion time scale. To this end, we plot the relative mobility, namely the ratio between the diffusion coefficient of the fillers and that of the polymers, $D_f/D_p$, in the bottom panel of Fig.\ref{fig:diff}. The relative mobility does not have a positive correlation with the stress at failure. At low loadings, relative mobility decreases while the stress at failure increases for all the filler sizes considered.  In the 15-25\% loading range, the SF and MF systems have comparable relative mobilities, but the SF system is remarkably tougher (see Fig. \ref{fig:stressstrain}). Moreover, the relative mobility does not have a monotonic dependence on loading and, at least for the SF in the loading range considered, it exhibits a clear minimum. Such a minimum corresponds to the maximum value for the stress at failure of the SF nanocomposite, which sets to 700 MPa in the 21-31\% loading range and is lower elsewhere.\\

\subsection*{Polymer-filler network}
\begin{figure}[th]
\includegraphics[width=0.48\textwidth]{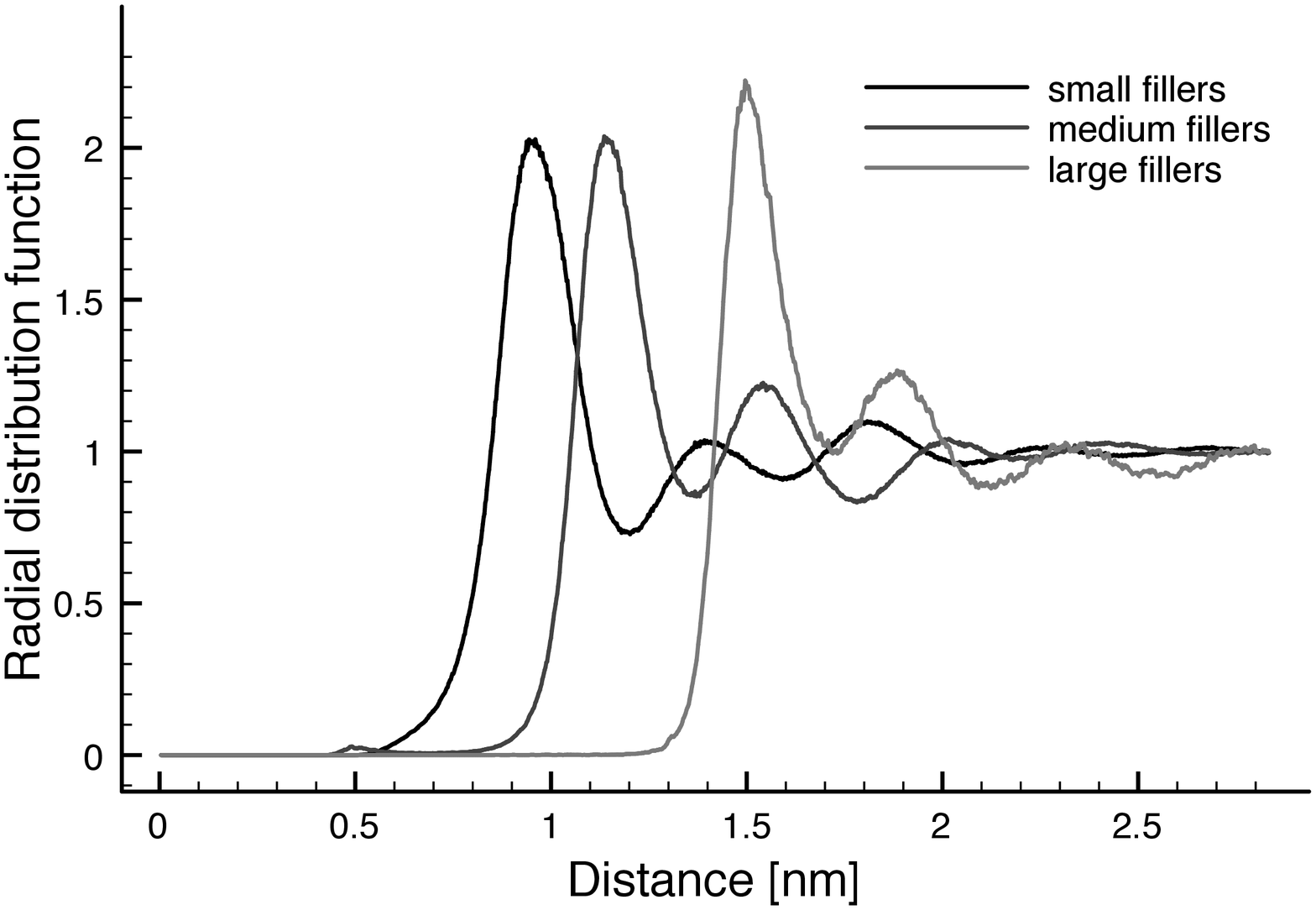}
\includegraphics[width=0.15\textwidth]{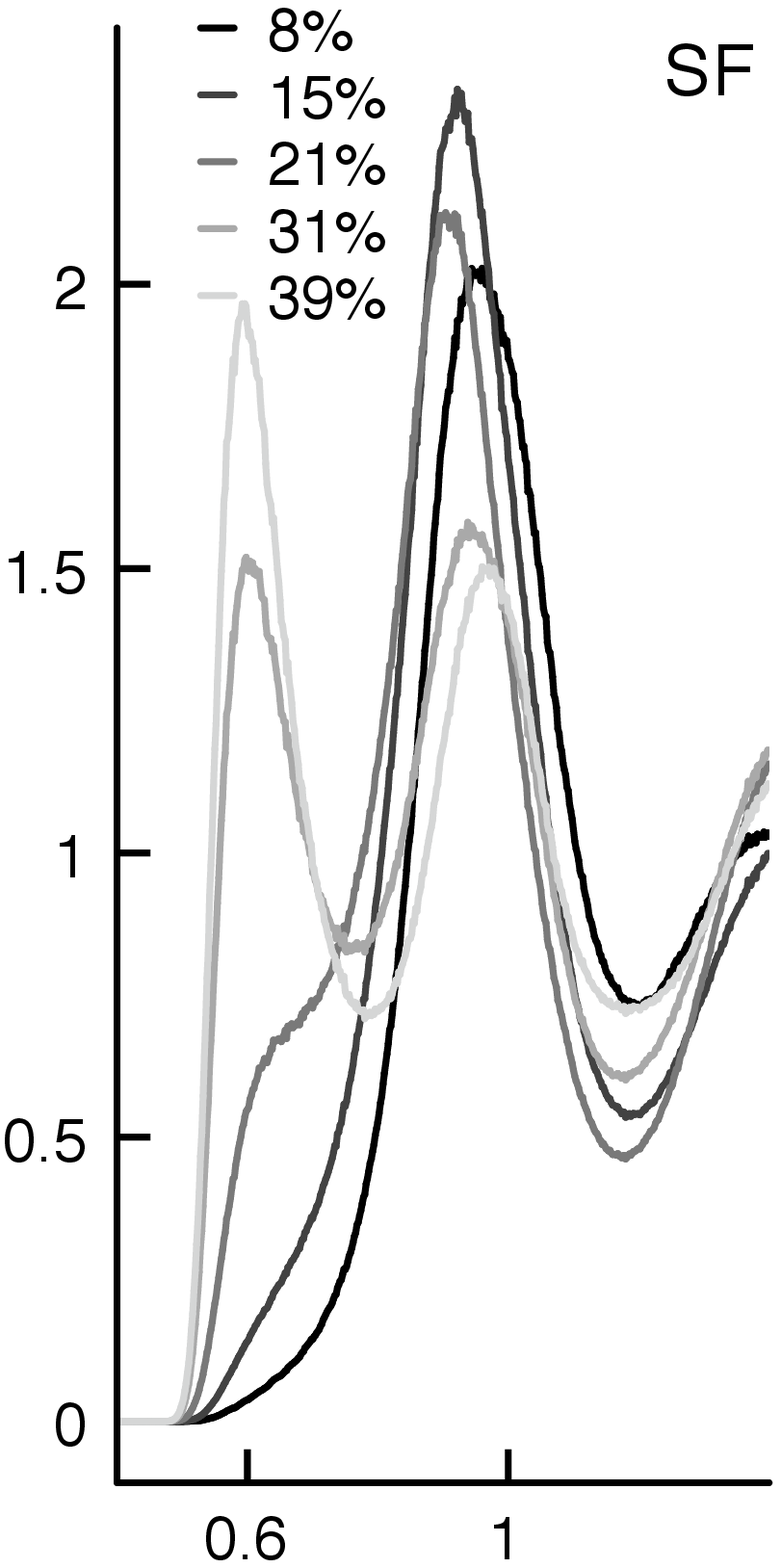}
\includegraphics[width=0.15\textwidth]{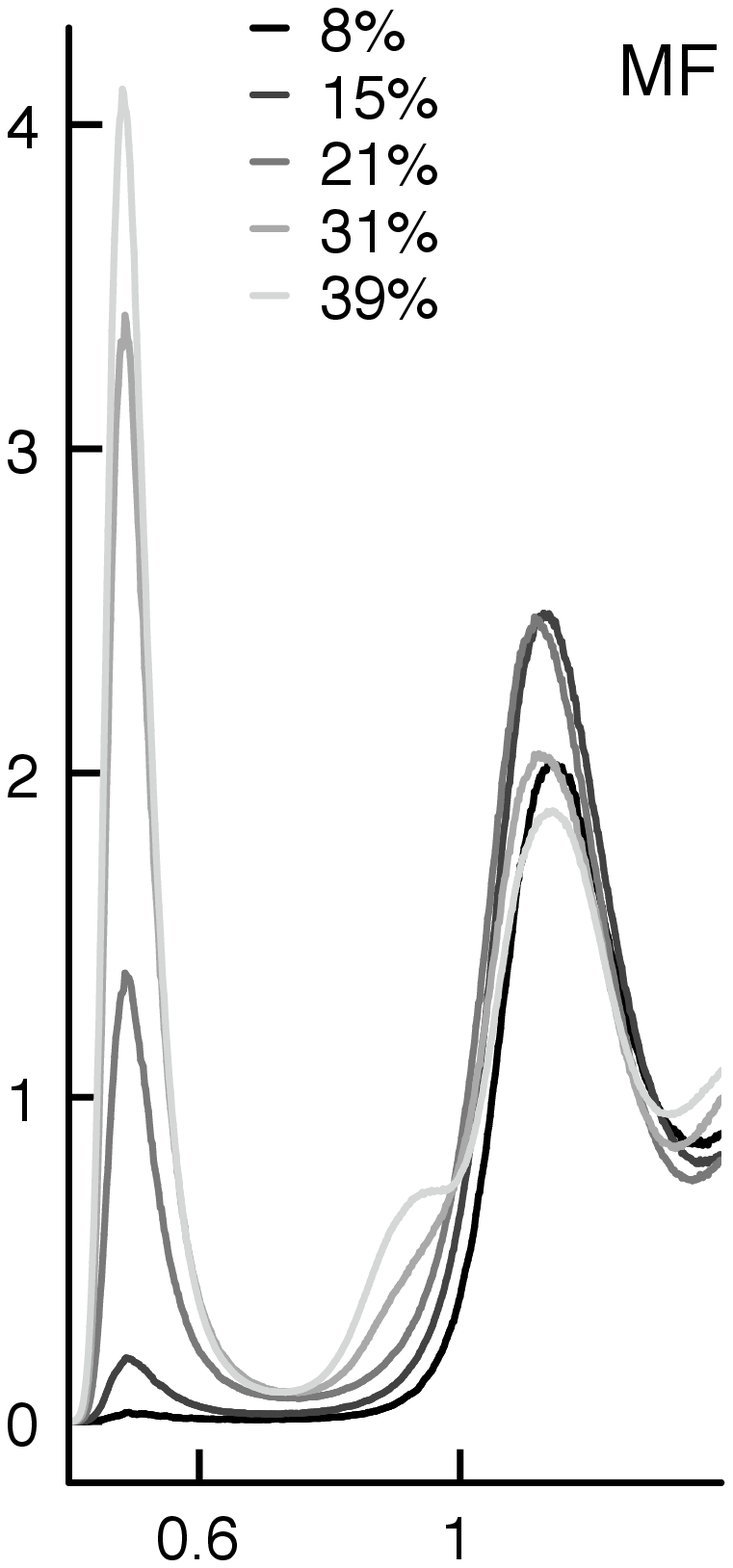}
\includegraphics[width=0.15\textwidth]{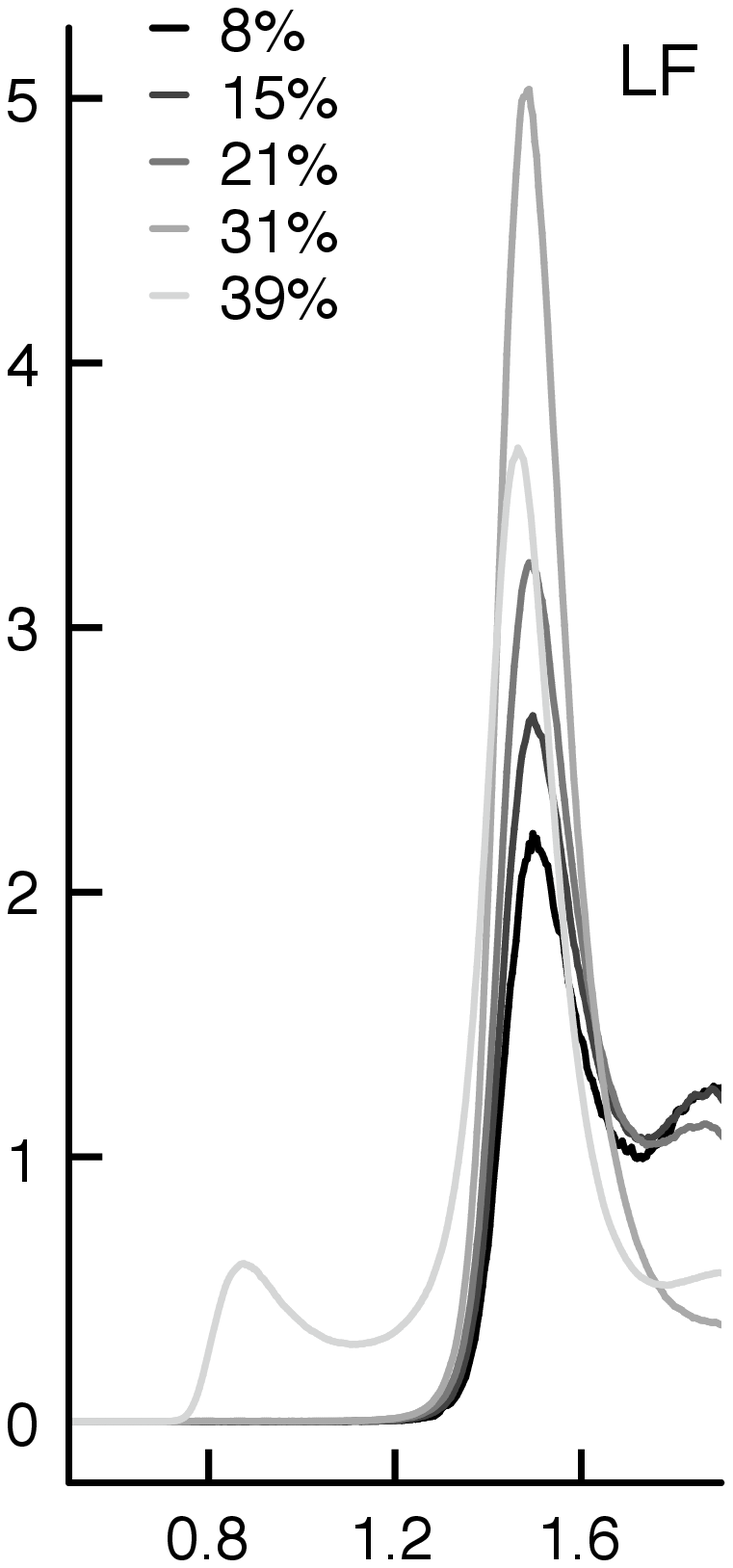}
\caption{\label{fig:rdf} Filler-filler radial distribution functions. In the top panel, the data are referred to three PNC containing 8\% loading of SF, MF or LF. In the bottom panel, the evolution of the first peaks of the radial distribution functions vs. loading is shown. Each panel is referred to a different bead size, SF, MF and LF, from left to right. The x and y axes report the same quantities as in the upper panel.}
\end{figure}

These results suggest to go back to structural sources of reinforcement, and in particular to an analysis of the structure of the polymer network as it is formed via the formation of strong polymer-filler contacts. Let us look at the filler-filler radial distribution functions. 
At 8\% loading, as shown in the top panel of Fig. \ref{fig:rdf}, no filler-filler contacts are formed in the matrix. The first peak of the RDF roughly corresponds, for all filler sizes considered, to bridging \cite{LiuPCCP11,Hooper05,Sen07} configurations. In such configurations, the fillers self-organize around polymer chains, in such a way that their first filler-filler neighbors are found only one polymer bead apart. This arrangement maximizes the strong polymer-filler contacts. It is then interesting to observe how RDF peaks evolve with loading. For small fillers, the bridging peak reaches a maximum at 15\% loading. Then, the peak corresponding to neighboring SF particles begins to be populated. At 31\% loading the two peaks are roughly equivalent and at 39\% the neighboring peak is the highest. This trend is correlated to our mechanical and dynamical indicators. Between 15\% and 31\% loading, in fact, the SF-loaded nanocomposites exhibit the largest stress at failure, while the filler mobility relative to the polymer matrix is suppressed. 31\% loading marks an inversion of the tendency for all of our mechanical, dynamical and structural indicators: the stress at failure decreases; the relative mobility of the SF increases;  the number of weak filler-filler contacts becomes comparable to the number of bridging configurations. \\
Despite the smaller number of MF in the systems, it is again at 15\% loading that the bridging peak reaches its maximum and then starts decreasing, in favor of the formation of the neighbor fillers peak. At 39\% a shoulder corresponding to second MF neighbors appears, as well. Medium fillers are thus much less efficient in creating bridging configurations, and the overall response to strain is weaker than in the SF case. As for the LF, the PNC is much less crowded and the bridging peak maximum sets at 31\% loading.\\
As a further structural indicator, we can count the number of different chains the nanofillers are, on average, in contact with. We thus consider multiple contacts between the same chain and a NP as a single bridging link, and then evaluate the average number of bridging links per filler. Fillers with a large surface area are expected to be more effective, namely to be able to create more bridging links than the fillers with small surface area. Still, if we normalize by $\sigma^2$, SF turn out to be the best at exploiting their surface area, namely $n^l_{SF}/\sigma^2_{SF} = 0.16 > n^l_{MF}/\sigma^2_{MF} = 0.11 > n^l_{LF}/\sigma^2_{LF} = 0.07 \AA^{-2}$ .

\section{Conclusion}

To conclude, we can thus rationalize two different contributions to mechanical reinforcement that make our small fillers more effective than the larger ones. The first reason is surface area, allowing for the formation of more, strongly attractive, polymer-chain bonds. In this respect, our results are consistent with Gersappe \cite{Gersappe02} and show that PNC containing fillers of different sizes, but sharing the same interfacial surface area, are still reinforced more effectively by smaller fillers. 
We have demonstrated further that the effectiveness of small fillers in reinforcing the matrix is associated with their ability to surround the polymer chains, maximizing bridging configurations. Such a structural configuration is correlated to a dynamical feature, namely the minimization of the relative mobility of the fillers with respect to the polymer matrix. This is in contrast with the hypothesis that the large mobility of the small fillers is responsible for the dissipation of mechanical energy, thus explaining their better reinforcement effect. Given the fast time scale of the simulated tensile tests, no dissipative effects are observed. \\

\section{Acknowledgements}

Authors acknowledge support by MatOx Ltd. Sakari R. Puisto 
and Niko Rostedt, from MatOx Ltd, and Chris Lowe and Bengt
Ingman, from Becker Industrial Coatings Ltd, are especially 
acknowledged for contributing with interesting discussions. 
This work has been supported in part by the Academy of Finland 
through its COMP CoE grant. MatOx Ltd acknowledges support by the Finnish
Funding Agency for Technology and Innovation, TEKES. Computing time at CSC is gratefully acknowledged.

\end{document}